# Fabrication of quantum emitters in aluminium nitride by Al–ion implantation and thermal annealing

E. Nieto Hernández[1,2], H.B. Yağcı[3,4], V. Pugliese[1,2], P. Aprà[1,2], J. K. Cannon[3,4], S. G. Bishop[3,4], J. Hadden[3,4], S. Ditalia Tchernij[1,2], Olivero[1,2], A.J. Bennett[3,4], J. Forneris[1,2]*

[1] Dipartimento di Fisica e centro inter-dipartimentale "NIS", Università di Torino, via Pietro Giuria 1, Torino, 10125, Italy
[2] Istituto Nazionale di Fisica Nucleare (INFN), Sezione di Torino, via Pietro Giuria 1, Torino, 10125 Italy
[3] School of Engineering, Cardiff University, Queen's Building, The Parade, Cardiff, CF24 3AA, United Kingdom
[4] Translational Research Hub, Cardiff University, Maindy Road, Cathays, Cardiff, CF24 4HQ, United Kingdom

**Abstract**

Single-photon emitters (SPEs) within wide-bandgap materials represent an appealing platform for the development of single-photon sources operating at room temperatures. Group III- nitrides have previously been shown to host efficient SPEs which are attributed to deep energy levels within the large bandgap of the material, in a way that is similar to extensively investigated colour centres in diamond. Anti-bunched emission from defect centres within gallium nitride (GaN) and aluminium nitride (AlN) have been recently demonstrated. While such emitters are particularly interesting due to the compatibility of III-nitrides with cleanroom processes, the nature of such defects and the optimal conditions for forming them are not fully understood. Here, we investigate Al implantation on a commercial AlN epilayer through subsequent steps of thermal annealing and confocal microscopy measurements. We observe a fluence-dependent increase in the density of the emitters, resulting in creation of ensembles at the maximum implantation fluence. Annealing at 600 °C results in the optimal yield in SPEs formation at the maximum fluence, while a significant reduction in SPE density is observed at lower fluences. These findings suggest that the mechanism of vacancy formation plays a key role in the creation of the emitters, and open new perspectives in the defect engineering of SPEs in solid state.

**Keywords**: single photon emitters, colour centres, III-nitrides, ion implantation

## 1 Introduction

Single photon emitters (SPEs) in wide-bandgap semiconductors are promising building blocks for quantum technologies, including quantum sensing, optical quantum computing and quantum communication[1-3]. Quantum emitters within solid-state host materials have steadily gained relevance in the last two decades. Alongside single photon emission in quantum dots[4], the experimental demonstration of anti-bunched emission from colour centres in diamond[5] led to a vibrant scientific field focused on characterization, manipulation, and fabrication of defect systems in wide-bandgap semiconductors[6-8].

With the advancements in ion implantation technology regarding deterministic single-ion doping and nanoscale placement precision[9-12], along with the substantial progresses in material synthesis and development in terms of controlled and selective chemical vapour deposition[13-15], it became possible to identify and develop optically-active defects with stable and efficient single photon emission, that in several instances is correlated to their (highly coherent) spin properties.

To date, a multitude of single photon emitters in wide band-gap semiconductors have been reported, both in the visible and in the infrared spectral regions[16-18]. The two most widespread materials in this field are diamond[19-23] and silicon carbide (SiC)[16, 24, 25]. Nonetheless, remarkable properties have been also demonstrated in other materials, such as silicon (Si)[26 - 28], gallium nitride (GaN)[17, 29, 30], hexagonal boron nitride (hBN)[31-33] and aluminium nitride (AlN)[34-36].

Aluminium nitride (AlN) is a wide-bandgap semiconductor ($E_g$ = 6.03 eV) with a refractive index of ~2.15 at $\lambda$ = 650 nm. It is well known and employed as a piezoelectric material, a durable ceramic, and the ideal buffer layer for GaN growth[37], making it an appealing semiconductor for the implementation of high-power electronics and next-generation photonics. Following theoretical studies suggesting the availability of optical dopants with spin properties similar to those of the diamond-based nitrogen-vacancy centre[38-40], the experimental demonstration of single-photon emission from native AlN luminescent defects was recently reported[29, 35, 36]. A subsequent study also showed the possibility to create luminescent extrinsic defects related via controlled laser-induced damage to the AlN crystal interface with a sapphire substrate[41]. However, the origin of these emitters is still not fully understood.

In this work, we inspect the manufacturability of single-photon emitters in AlN thin films by means of Al ion implantation and subsequent thermal annealing, in order to identify a reliable protocol to produce single-photon emitters from intrinsic point defects. We use Al ion implantation to promote the formation of lattice vacancies without introducing extrinsic defect complexes (i.e. defects related to the introduction of foreign chemical species) into the material. We investigate the correlation between the implantation fluence and the overall photoluminescence (PL) emission intensity from the implanted crystal, leaving out of the study the discrimination between the implanted quantum emitters and the previously present in the material. Furthermore, we assess the role of the subsequent thermal annealing to both activate individual luminescent centres and reduce the radiation-induced background emission. Finally, by means of single-photon confocal photoluminescence characterization, we show that vacancy-related quantum light emitters in AlN can be reliably fabricated by means of ion implantation combined with suitable thermal processing.

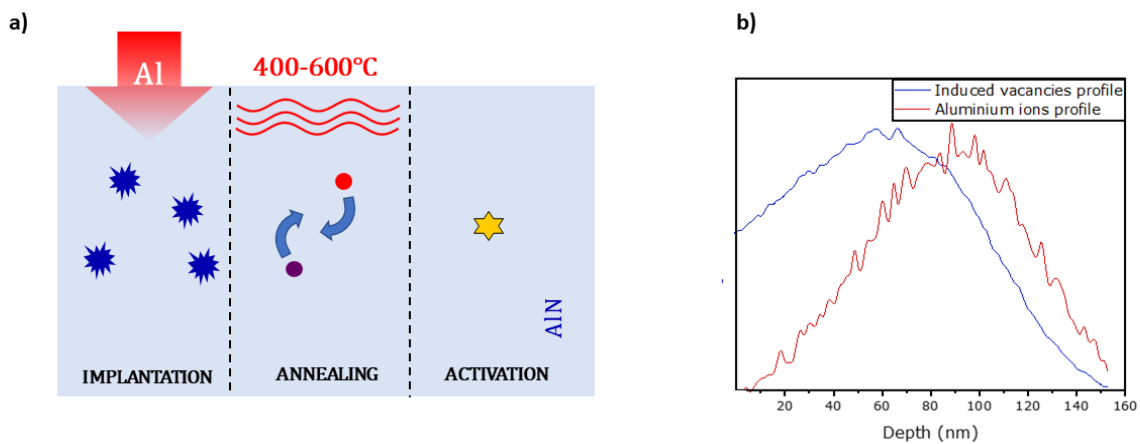

*Figure 1*. Schematic overview of the experimental work. *a)* Scheme of the fabrication protocol adopted in the present work. Following ion implantation, the annealing-induced thermal diffusion of vacancies and/or impurities within AlN results in the formation of stable, optically-active emitters within AlN. *b)* Vacancy density profile (blue) and the implanted Al ion distribution (red) associated with 60 keV Al implantation, calculated via SRIM Monte Carlo code[42].

## 2 Experimental methods

The experiments were performed on a 1 µm thick Metal-Organic Chemical Vapour Deposition (MOCVD)-grown AlN epilayer on sapphire purchased from Dowa Electronics Materials Co. The material is a wurtzite-type single crystal and contains native optically-active defects dispersed with low density[36]. The overall procedure adopted for the fabrication and characterization of the samples is highlighted in **Figure 1**. In particular, **Fig. 1a** shows the fabrication process performed on the substrate. The implantation of Al ions induces the formation of lattice vacancies, whose diffusion and recombination are promoted by means of thermal annealing. The resulting effect is the formation of optically-active colour centres. Specifically, the sample was implanted with 60 keV Al⁻ ions with the recently established multi-elemental ion source of the

Solid State Physics laboratories of the University of Torino. Several circular regions of ø~1 mm were implanted at different fluences in the $10^{12}$-$10^{14}$ cm$^{-2}$ range by means of a movable collimating mask. **Fig. 1b** shows the ion implantation profile for the considered energy as a function of the sample depth along with the corresponding vacancy density profile, suggesting that the implantation-induced colour centres are formed within 100 nm from the surface. Multiple subsequent post-implantation annealing treatments were performed using a tubular oven at 400 °C and subsequent multiple 30 min processes at 600 °C in $N_2$ atmosphere.

The sample was characterised by means of PL confocal microscopy between the annealing treatments, in order to assess the effects of ion implantation and annealing on the formation of optically active centres. The implanted regions were marked by means of high-power laser milling to enable the investigation of the same photoluminescent regions after each of the subsequent treatments performed on the sample. The PL analysis was performed using a custom fibre-coupled single-photon sensitive microscope equipped with a 100x dry objective (0.9 NA). Optical excitation was supplied by a 520 nm laser diode. In all measurements a set of spectral filters defined a detection spectral window in the 550-650 nm range, thus ensuring the removal of the background originating from the chromium line of sapphire[43] and an abrupt cut off in the emitter spectra at the edges of this range. Single-photon emission qualification was performed on isolated photoluminescent spots using a Hanbury-Brown & Twiss (HBT) interferometer implemented by a multimode fibre-fused 50:50 beamsplitter coupled to two independent single photon avalanche photodiodes (SPADs). The assessment of the density of created centres was performed by comparing the number of isolated spots found in the implanted regions by PL mapping[44], with respect to a reference pristine region. Finally, spectral characterization at room temperature was performed on all of the considered regions using a Princeton Instruments PIXIS spectrometer.

## 3 Results and Discussion

**Emitter formation upon ion implantation and annealing.** The effects of Al ion implantation on the formation of colour centres in AlN were investigated by PL confocal microscopy mapping as a function of implantation fluence and post-implantation annealing temperature. **Figure 2** shows the PL maps (20x20 μm$^2$ area, 2 mW excitation power) of the unimplanted region and of the regions implanted at $1 \times 10^{12}$ cm$^{-2}$, $1 \times 10^{13}$ cm$^{-2}$ and $1 \times 10^{14}$ cm$^{-2}$ fluences after each of the above-mentioned thermal treatments. The maps acquired prior to any subsequent annealing (condition referred to as "as implanted" in the following) are presented in the top row of **Fig. 2a-d**. The pristine region (first column) exhibits a low density of single emitters (~0.075 μm$^{-2}$), corresponding to individual bright spots (~50 kcps emission intensity) with respect to the background originating from the surrounding region (~2 kcps). The single-photon emission from individual spots was verified by HBT interferometry on a set of ~10 emitters for all each region.

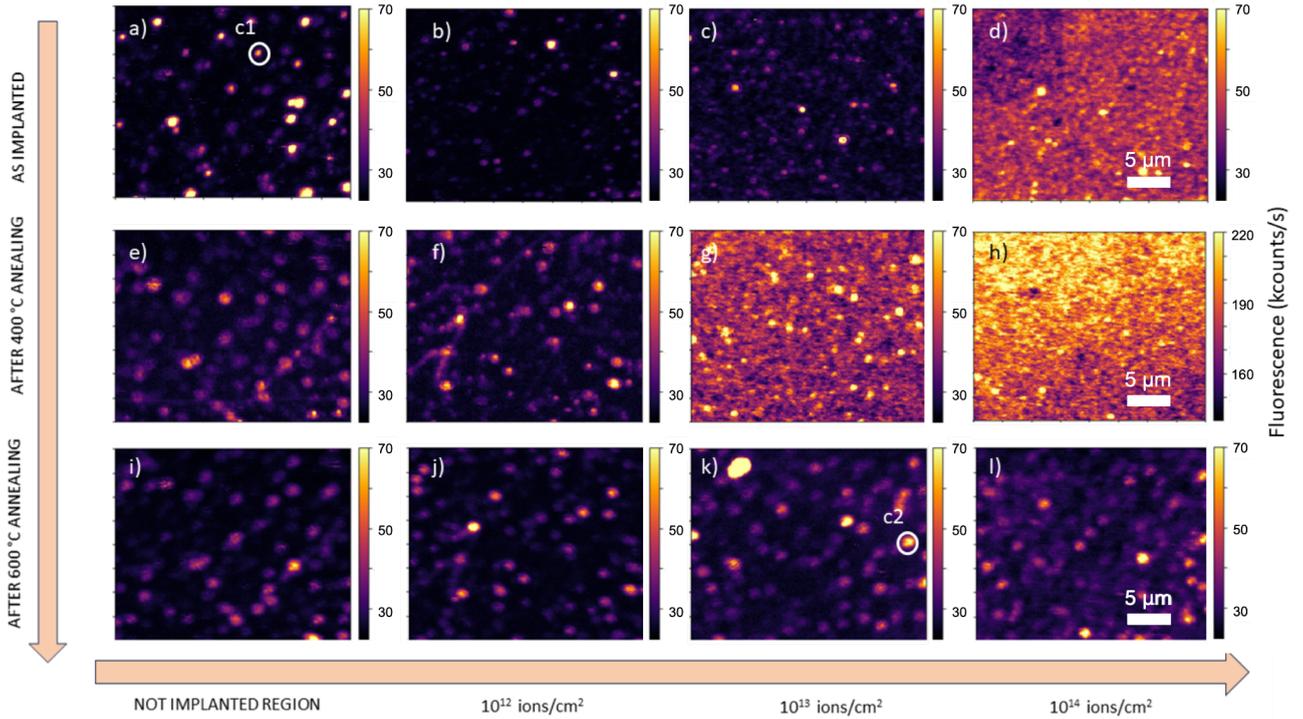

*Figure 2.* PL confocal mapping of the same processed regions (20x20 µm² scan area) upon different thermal annealing steps under 520 nm laser excitation and collection window between 550-650 nm. The colour scale encodes the range 0-70 kcps for all the considered PL scans, except for the region implanted at 1x10¹⁴ cm⁻² fluence and annealed at 400 °C, due to the high overall emission intensity of the latter.

The characterization of an exemplary emitter (corresponding to the emission spot circled in white and labelled as "C1" in **Fig. 2a**) is shown in **Fig. 3**. The second-order auto-correlation histogram acquired via HBT interferometry at different optical excitation powers (Fig. 3a) highlight the occurrence of non-classical emission originating from an individual source (i.e. $g^{(2)}(t=0)<0.5$). The curves were fitted according to a model based on a three-level system that allows for the presence of a metastable shelving state[2,45]:

$$g^{(2)}(t)=1-(1+a)\cdot\exp(-|t|\cdot\lambda_1)+a\cdot\exp(-|t|\cdot\lambda_2) \tag{1}$$

Under vanishing optical excitation power, the decay rate through the shelving state of the three level system is negligible with respect to that of the radiative transition. The excited state lifetime was estimated as (6.7±0.2) ns by a linear regression of the $\lambda_1$ parameter versus the optical excitation power (Fig. 3b). The emission intensity at saturation was determined (**Fig. 3c**) to be (110±5) kcps at an optical power of (3.6±0.3) mW. The spectral signature of the emitter consisted of a broad band in the 550-650 nm, probably extending outside the range pass-band of the filtering optics.

The region implanted at 1x10¹² cm⁻² fluence (**Fig.2b**) did not exhibit any significant increase in the density of emitters (~0.08 µm⁻²) with respect to the unimplanted region. In contrast, the map acquired from the region implanted at 1x10¹³ cm⁻² (**Fig. 2c**) shows an increase in the emitter density (~0.37 µm⁻²), each displaying comparable photon count rates (i.e. 60 kcps under 1 mW excitation optical power) with respect to the pristine region. This observation is indicative of the fact that ion implantation is responsible for the formation of optically active defects[45], as supported by the further increase in the overall PL intensity (~100 kcps) observed upon ion implantation at 1x10¹⁴ cm⁻²(**Fig. 2d**). In this latter case, the substantial PL increase prevents the identification of individual emitters.

Following the PL characterization of the as-implanted sample, the above-listed thermal processing steps were carried out. **Figs. 2d-h** and **Figs. 2i-l** report the PL maps (20x20 µm² area) of the unimplanted region and of the regions implanted at 1x10¹²-1x10¹⁴ cm⁻² fluences after 400 °C and 600 °C treatments, respectively. The same trend of the emitter density on the ion fluence, as highlighted prior to the thermal processing for all the irradiation conditions, was observed. A first annealing at 400 °C did not significantly modify the emitter density in the pristine sample (**Fig.2e**), while it resulted in an increase (0.1–0.2 µm²) in the areal density of emitters in the regions implanted at 1x10$^{12}$ (**Fig.2f**) and 1x10$^{13}$ (**Fig. 2g**) cm$^{-2}$ fluences, although the latter was accompanied by a noticeable increase in the background PL emission. Notably, none of the native centres disappeared as a result of the 400 °C process.

The region implanted at 1x10$^{14}$ cm$^{-2}$ fluence (**Fig. 2h**) exhibited a three-fold increase in the PL intensity, indicating the formation of a large density of optically active intrinsic defects, whose overall emission prevented in this case the identification of individual emitters. The subsequent treatment at 600 °C did not alter the initial density of emitters in the pristine region (**Fig. 2i**) with respect to the previous treatments. Conversely, in the case of the second thermal process the 1x10$^{13}$-1x10$^{14}$ cm$^{-2}$ ion fluences (**Figs. 2j-l**) this higher temperature process suppressed the background PL emission that was observed after the 400 °C treatment, thus resulting in a distribution of single emitters with a higher visibility and with a 0.20-0.25 µm$^{-2}$ areal density, respectively.

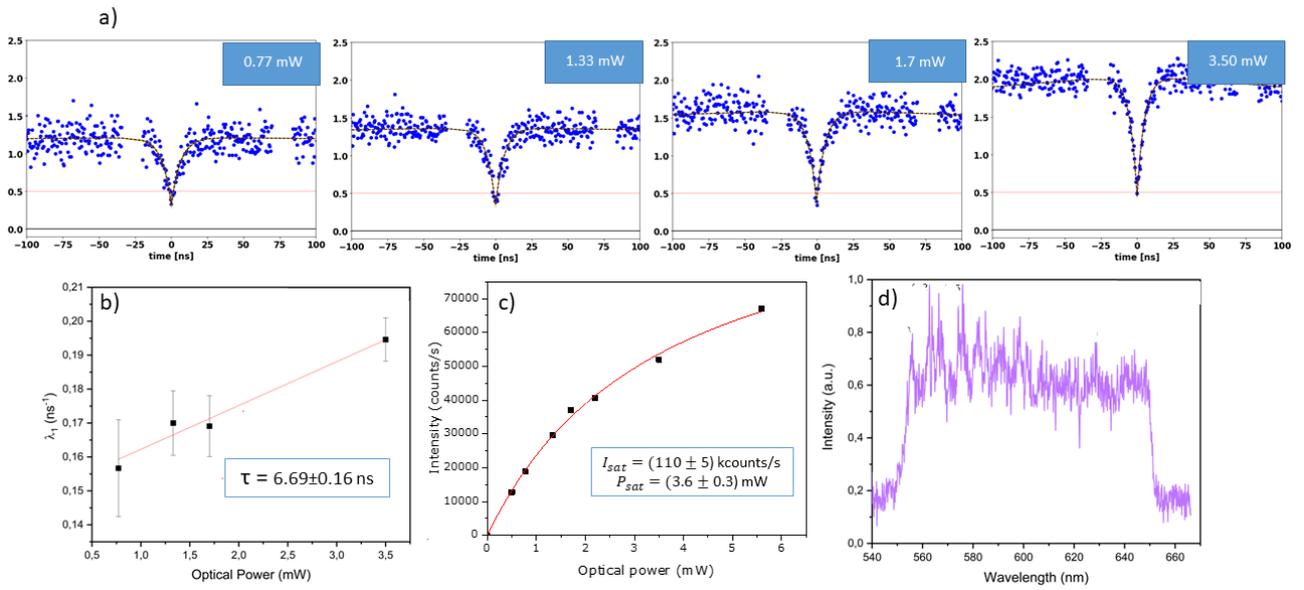

*Figure 3. Single-photon emission parameters from an individual centre in the unimplanted region prior to any thermal processing. The emitter is circled in white and labelled as "C1" in Fig. 2a: **a)** Second order auto-correlation chronograms. The missing datapoints corresponds to the backflash peaks of the detection system, which were removed to enhance the readability of the graph. **b)** Linear regression of the $\lambda_1$ parameter as a function of the excitation power allowing the estimation of the reported spontaneous emission lifetime value. **c)** Saturation curve for the emitter, after background subtraction. **d)** PL emission spectrum.*

The nature of single-photon emitters in ion implanted and annealed AlN was investigated with the same characterization procedure adopted for the unprocessed substrate in **Fig. 3**. **Figure 4** shows the single-photon emission properties of an individual emitter located in the region processed by Al implantation at 10$^{13}$ cm$^{-2}$ fluence and 600 °C annealing (the centre is labelled as "C2" in **Fig. 2k**). The measurement of g$^{(2)}$(t=0)<0.5 in the HBT chronograms confirmed the occurrence of single-photon emission (**Fig. 4a**), which was modelled with the same three-level system model adopted for the "C1" centre (see Eq. 1). The analysis resulted in an estimation of the radiative lifetime of the "C2" centre as (3.39 ± 0.2) ns. The emission intensity and excitation power at saturation parameters (**Fig. 4c**) were determined to be (119±9) kcps and (0.28±0.09) mW, respectively. While the emission intensity at saturation is comparable, the centre's lifetime and excitation power in saturation conditions are somewhat different with respect to the "C1" centre. This could be explained by considering the different spectral signature of the "C2" centre (**Fig. 4d**), which exhibits a large band with a broad emission peak around 570 nm. Such a feature is not observed for the "C1" emitter. The two emitters could therefore be attributed to different defective complexes, both being compatible with what was previously reported in the scientific literature for unimplanted AlN[36]. A further discussion on the spectral variability of the emitters investigated in this sample is reported in the following section.

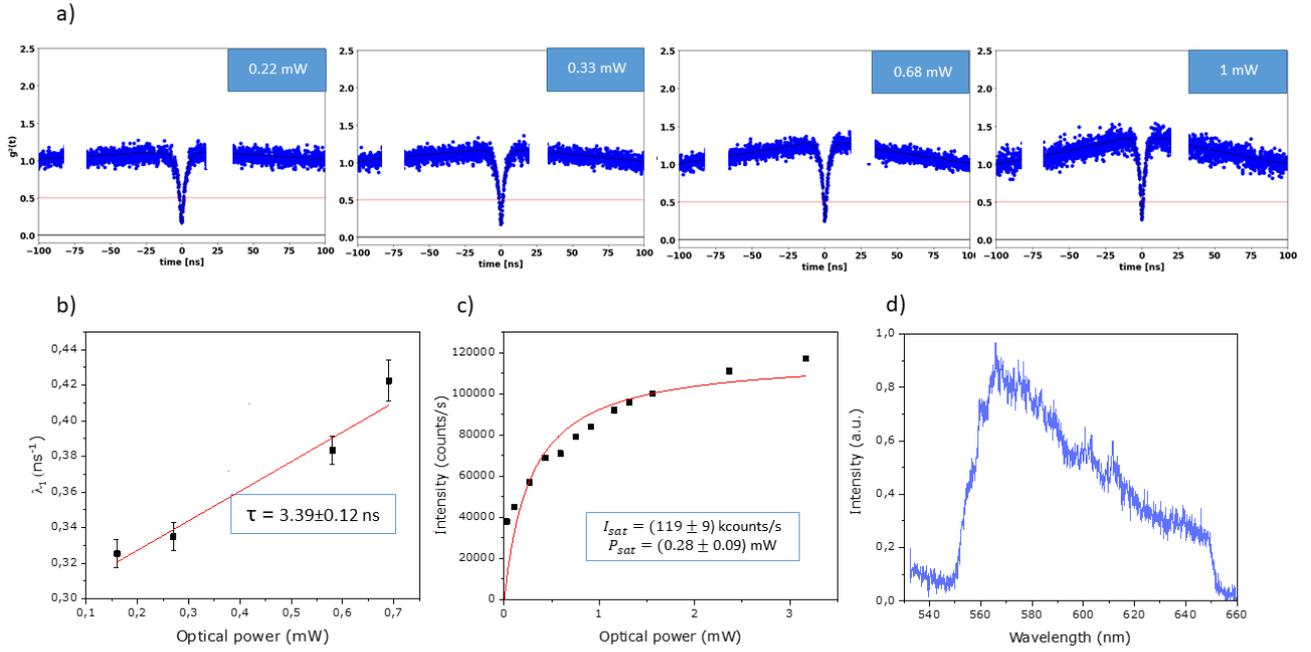

*Figure 4.* Single-photon emission parameters from an individual centre in the region implanted at $10^{13}$ cm$^{-2}$ fluence after the 600°C thermal processing. The emitter is circled in white and labelled as "C2" in Fig. 2a: a) Second order correlation measurements and lifetime extracted from the power dependence of the $\lambda_1$ fitting parameter. The missing data points corresponds to the backflash peaks of the detection system, which were removed to enhance the readability of the graph. b) Linear regression of the $\lambda_1$ parameter as a function of the excitation power. c) Saturation curve of the emitter, after background subtraction. d) PL spectrum with spectral filtering to remove light outside the 550-650 nm range.

In order to quantify the occurrence of single emitters following each fabrication step, the respective areal density is shown in **Fig. 5** for the unimplanted sample, and the regions implanted at $1\times10^{12}$-$1\times10^{14}$ cm$^{-2}$. The maximal density of emitters (0.4 µm$^{-2}$) is found in the $1\times10^{13}$ cm$^{-2}$ ion fluence prior to any annealing treatment. The 400 °C annealing results in a moderate increase in the density (up to 0.2 µm$^{-2}$) for the $1\times10^{12}$ cm$^{-2}$ fluence only, while inducing a significant background PL at higher ion fluences. Finally, the 600 °C results in a slight decrease in the density (0.15 µm$^{-2}$) for $1\times10^{12}$ cm$^{-2}$ ion fluence with respect to the previous treatment, and a significant decrease (0.2 µm$^{-2}$) for the $1\times10^{13}$ cm$^{-2}$ ion fluence with respect to the unprocessed sample. Ultimately, the 600 °C annealing with the highest implantation fluence ($1\times10^{14}$ cm$^{-2}$) results in the formation of individual emitters. We note that for several combinations of ion fluence and annealing conditions we are unable to quantify the density of individual emitters because the high levels of PL coming from the whole regions, suggesting the formation of ensembles of centres not individually resolvable. These combinations have not been included in **Figure 5**.

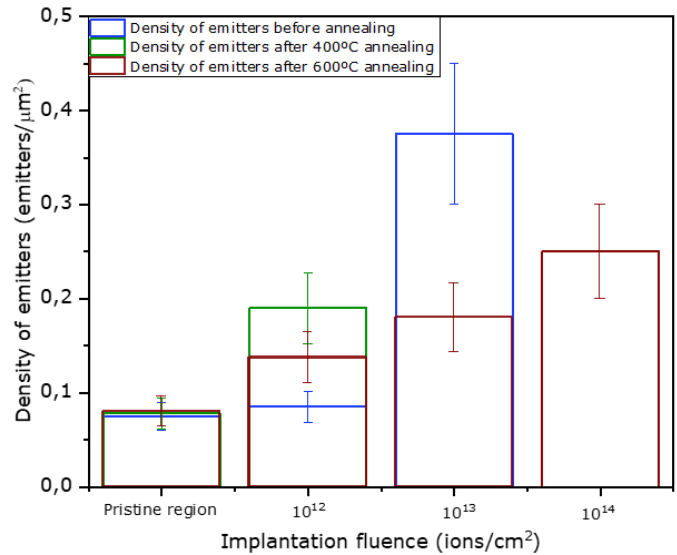

*Figure 5.* Boxchart representation of the areal distribution of single-photon emitters in AlN implanted at different Al–fluences following different thermal processes, namely: untreated, 400 °C annealing, and 600 °C annealing.

**Spectral features of AlN emitters**. The spectra reported in **Figs. 3d, 4d** are in good agreement with what has been reported in literature, but not yet attributed to any specific defective configuration [36]. **Figure 6** shows a statistical analysis of the spectral signature of a set of 58 PL spectra acquired from individual emitters. All the emitters are characterised by large (i.e.>20 nm) spectral bands as in the case of the exemplary emitters shown in **Figs. 3-4**, which is centred around a peak wavelength. **Figure 6a** shows the statistical distribution of the central peak wavelength as a function of the annealing temperature, independently of the ion implantation fluence at which the region containing each individual emitter was processed. Conversely, **Fig. 6b** highlights the peak wavelength occurrence as a function of the implantation fluence, independently of the post-implantation temperature treatment. In this latter case, no centres from the region implanted at $1\times10^{13}$ cm$^{-2}$ and $1\times10^{14}$ cm$^{-2}$ and subsequently annealed at 400 °C could be identified due to the intense background PL. In all cases, the peak central wavelength displayed a large variability in the 560-620 nm range, with the majority of the centres emitting between 570 nm and 600 nm. No specific trend in the central wavelength distribution was observed neither as a function of the implantation fluence nor the annealing temperature, besides the lack of emission centred at 620 nm following any of the considered thermal processes.

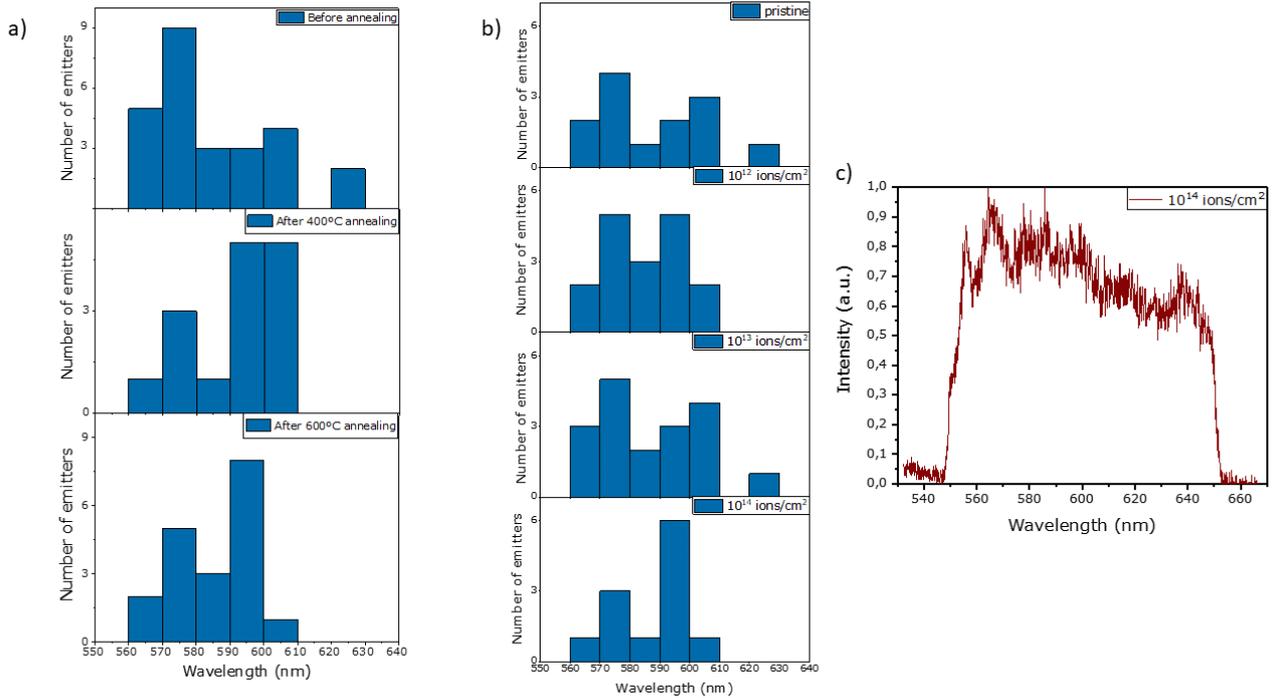

*Figure 6.: Statistical distribution of the central wavelength of the main PL emission peaks from a set of individual emitters in AlN. **a)** distribution of the emitters at different annealing temperatures; **b)** distribution for different ion implantation fluences. **c)** PL spectra of the background PL emission acquired from the region implanted at the highest ion fluence ($1\times10^{14}$ cm$^{-2}$) and annealed at 400°C, corresponding to the PL map in **Fig. 2h**.*

The variability in the distribution of the central peak wavelengths observed in the considered emitters might be indicative of the presence of different classes of emitters, or alternatively to local lattice effects on the electronic states of the point defects. This latter interpretation could be justified by the piezoelectric properties of AlN, as well as by local strains at the interface between the 1 μm thick AlN layer and the underlying sapphire substrate. To provide further context, a spectral analysis of the sample ensemble emission acquired from the region implanted at $1\times10^{14}$ cm-2 ion fluence and processed by 400 °C annealing. In this case, corresponding to the PL map shown in **Fig. 2h**, the broad PL spectra in the 560-640 nm range are reminiscent of the individual PL peaks observed at the single-photon emitter level, suggesting that the ensemble emission is a convolution of the multiple spectral components recorded in the histograms in **Fig. 6a-b**. This result confirms the role of ion implantation in the formation of optically-active intrinsic defects in AlN. The observation of ensemble emission with increasing intensity at the highest considered ion fluences further corroborates this interpretation since the overall PL emission increases with the density of ion-induced radiation damage in the host crystal[23].

Finally, an additional insight in the role of the thermal annealing was obtained by performing an analysis on an individual emitter located in the unimplanted region (centre "C3" in **Fig.7a**), following each thermal treatment. The "C3" centre exhibited a broad emission band centred at ~590 nm, which was left unchanged by the subsequent annealing steps. This observation indicates that the isolated defect was still optically active after annealing at temperatures as high as 600 °C and that its emission spectrum was not altered by the thermal process.

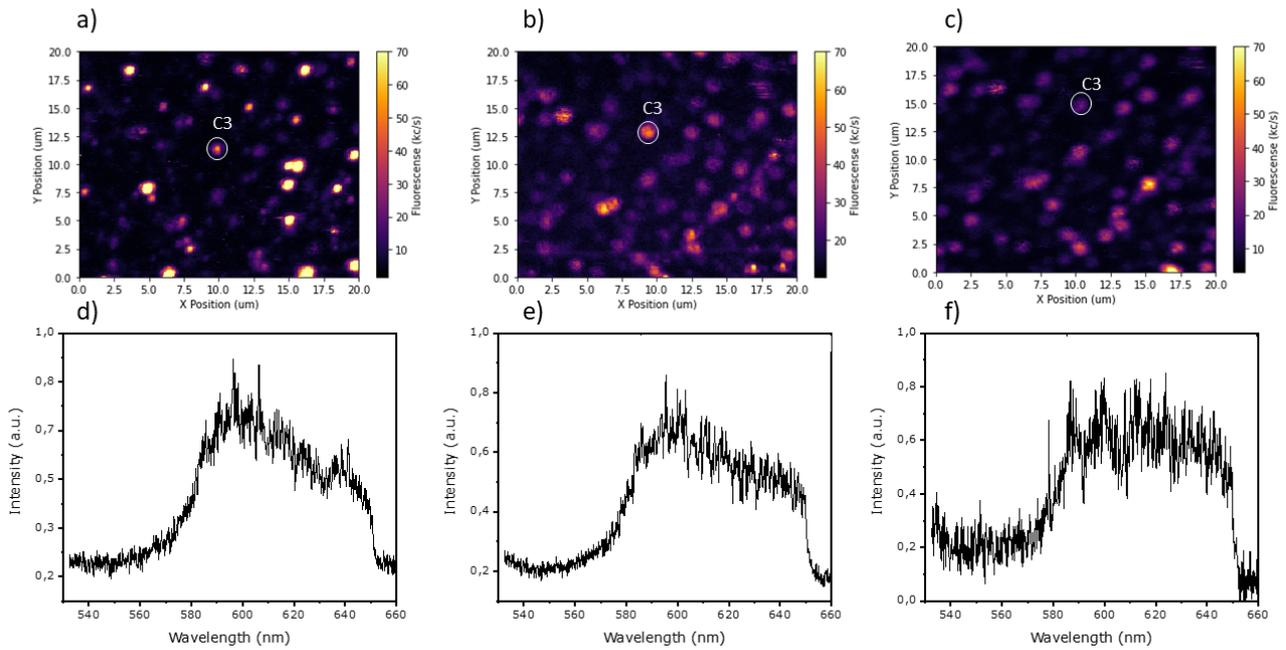

*Figure 7.* Spectral comparison of the same single emitter (labelled as "C3" in the PL maps isolated in the unimplanted region after each post-implantation process): a) pristine material; b) following 400 °C thermal annealing; c) following 600 °C thermal annealing.

## 4 Conclusions

In this paper we performed a systematic investigation on the role of ion implantation in the formation of single-photon emitting intrinsic defects in AlN. We demonstrated that the implantation of 60 keV Al⁻ ions increases the density of individual colour centres and the background emission intensity was correlated with the ion fluence. The study of which of these optically active quantum emitters corresponded to an already present centre and which to an implanted one was out of the scope of this work but would require a specific investigation of each emitter level to separately analyse the properties of the implanted defects. The highest formation yield was achieved upon irradiation at $1 \times 10^{13}$ cm$^{-2}$ fluence, without any subsequent thermal treatment of the sample. Higher ion fluences resulted in the formation of emitter ensembles that resulted in a limited capability to isolate single colour centres. A spectral analysis of the emitters revealed that all the point defects analysed in the ion implantation region have similar features with respect to those found in pristine AlN, thus highlighting that Al- ion implantation results in the formation of intrinsic defects. The structural and spectral stability of individual colour centres following thermal processes up to 600 °C was also demonstrated. These results show a viable pathway, based on industry-compatible ion implantation technique that in perspective can be potentially scaled to the single-ion delivery level[9,11], for the fabrication of quantum emitters in chip-integrable AlN platform.

## Acknowledgments

This work was supported by the following research projects:
IR-HPHT, funded by the Italian Ministry of University and Research DM 737/2012 within the National Programme for Research (PNR); "Training on LASer fabrication and ION implantation of DEFects as quantum emitters" (LasIonDef) project funded by the European Research Council under the "Marie Skłodowska-Curie Innovative Training Networks" program; "Departments of Excellence" (L. 232/2016), funded by [13] the Italian Ministry of Education, University and Research (MIUR); "Ex-post funding of research – 2021" funded by Compagnia di San Paolo.
The projects 20IND05 (QADeT) and 20FUN05 (SEQUME) and 20FUN02 (PoLight) leading to this publication have received funding from the EMPIR programme co-financed by the Participating States and from the European Union's Horizon 2020 research and innovation programme.

P.O. gratefully acknowledges the support of "QuantDia" project funded by the Italian Ministry for Instruction, University and Research within the "FISR 2019" program.

Cardiff University acknowledges financial support provided by UK's EPSRC via Grant No. EP/T017813/1 and EP/03982X/1 and the European Union's H2020 Marie Curie ITN project LasIonDef (GA No. 956387), and the the Sêr Cymru National Research Network in Advanced Engineering and Materials.

**References**

[1] D. D. Awschalom et al., "Quantum technologies with optically interfaced solid-state spins" Nature Photonics **12**, 516, (2018).

[2] C. Bradac et al., "Quantum nanophotonics with group IV defects in diamond" Nature Communications **10**, 5625 (2019).

[3] Rembold et al., "Introduction to quantum optimal control for quantum sensing with nitrogen-vacancy centers in diamond" AVS Quantum Science **2** 024701 (2020).

[4] Michler et al., "A Quantum Dot Single-Photon Turnstile Device" Science **290** 2282 (2000).

[5] C. Kurtsiefer et al., "Stable Solid-State Source of Single Photons" Physical Review Letters **85** 290 (2000).

[6] M. Radulaski et al., "Scalable Quantum Photonics with Single Color Centers in Silicon Carbide" Nano Letters **17** 1782 (2017).

[7] S. A. Momenzadeh et al., "Nanoengineered Diamond Waveguide as a Robust Bright Platform for Nanomagnetometry Using Shallow Nitrogen Vacancy Centers" Nano Letters **15** 165 (2014).

[8] X.-J. Wang et al., "Laser Writing of Color Centers" Laser Photonics Reviews **16** 2100029 (2021).

[9] M. Lesik et al., "Maskless and targeted creation of arrays of colour centres in diamond using focused ion beam technology" Phys Status Solidi A **210** 2055 (2013).

[10] T. Lühmann et al., "Coulomb-driven single defect engineering for scalable qubits and spin sensors in diamond" Nature Communications **10** 4956 (2019).

[11] J. L. Pacheco et al., "Ion implantation for deterministic single atom devices" Review of Scientific Instruments **88** 123301 (2017).

[12] Y. Zhou et al., "Direct writing of single germanium vacancy center arrays in diamond" New J. Phys. **20** 125004 (2018).

[13] L. Hussey et al. "Direct Observation of the Polarity Control Mechanism in Aluminum Nitride Grown on Sapphire by Aberration Corrected Scanning Transmission Electron Microscopy" Microscopy and Microanalysis **20** 162 (2014).

[14] R. Boichot et al. "Epitaxial and polycrystalline growth of AlN by high temperature CVD: Experimental results and simulation" Surface and Coatings Technology **205** 1294 (2010).

[15] D. Alden et al., "Fabrication and structural properties of AlN submicron periodic lateral polar structures and waveguides for UV-C applications" Applied Physics Letters **108** 261106 (2016).

[16] J. Wang et al., "Bright room temperature single photon source at telecom range in cubic silicon carbide" Nature Communications **9** 4106 (2018).

[17] Y. Zhou et al., "Room temperature solid-state quantum emitters in the telecom range" Science Advances **4** eaar3580 (2018).

[18] A. Senichev et al., "Room-temperature single-photon emitters in silicon nitride" Sci. Adv. **7** eabj0627 (2021).

[19] C. Wang et al., "Single photon emission from SiV centres in diamond produced by ion implantation" Journal of Physics B: Atomic, Molecular and Optical Physics **39** 37 (2005).

[20] T. Iwasaki et al., "Germanium-Vacancy Single Color Centers in Diamond" Scientific Reports **5** 12882 (2015).

[21] I. Aharonovich, S. Castelletto, B. C. Johnson, J. C. McCallum, D. A. Simpson, A. D. Greentree y S. Prawer, "Chromium single-photon emitters in diamond fabricated by ion implantation" Physical Review B **81** 121201 (2010).

[22] T. Gaebel et al., "Stable single-photon source in the near infrared" New Journal of Physics **6** 98 (2004).

[23] S. D. Tchernij et al., "Single-Photon-Emitting Optical Centers in Diamond Fabricated upon Sn Implantation" ACS Photonics **4** 2580 (2017).

[24] M. Widmann et al., "Coherent control of single spins in silicon carbide at room temperature" Nature Materials, **14** 164 (2014).

[25] S. Castelletto et al., "A silicon carbide room-temperature single-photon source" Nature Materials **13** 151 (2013).

[26] M. Hollenbach et al., "Engineering telecom single-photon emitters in silicon for scalable quantum photonics" Optics Express **28** 26111 (2020).
[27] A. Durand et al., "Broad Diversity of Near-Infrared Single-Photon Emitters in Silicon" Physical Review Letters **126** 083682 (2021).
[28] Y. Baron et al., "Detection of Single W-Centers in Silicon" ACS Photonics **9** 2337 (2022).
[29] Y. Xue et al., "Bright room temperature near-infrared single-photon emission from single point defects in the AlGaN film" Applied Physics Letters **118** 131103 (2021).
[30] A. M. Berhane et al., "Bright Room-Temperature Single-Photon Emission from Defects in Gallium Nitride" Advanced Materials **29** 1605092 (2017).
[31] G. Grosso et al., "Tunable and high-purity room temperature single-photon emission from atomic defects in hexagonal boron nitride" Nature Communications **8** 705 (2017).
[32] N. V. Proscia et al., "Near-deterministic activation of room-temperature quantum emitters in hexagonal boron nitride" Optica **5** 1128 (2018).
[33] Khatri et al., "Optical Gating of Photoluminescence from Color Centers in Hexagonal Boron Nitride" Nano Letters **20** 4256 (2020).
[34] T.-J. Lu et al., "Bright High-Purity Quantum Emitters in Aluminum Nitride Integrated Photonics" ACS Photonics **7** 2650 (2020).
[35] Y. Xue et al., "Single-Photon Emission from Point Defects in Aluminum Nitride Films" The Journal of Physical Chemistry Letters **11** 2689 (2020).
[36] S. G. Bishop et al., "Room-Temperature Quantum Emitter in Aluminum Nitride" ACS Photonics **7** 1636 (2020).
[37] C. Xiong et al., "Aluminum nitride as a new material for chip-scale optomechanics and nonlinear optics" New Journal of Physics **14** 095014 (2012).
[38] J. R. Weber et al., "Quantum computing with defects" Proceedings of the National Academy of Sciences **107** 8513 (2010).
[39] Y. Tu et al., "A paramagnetic neutral VAlON center in wurtzite AlN for spin qubit application" Applied Physics Letters **103** 072103 (2013).
[40] H. Seo et al., "Design of defect spins in piezoelectric aluminum nitride for solid-state hybrid quantum technologies" Scientific Reports **6** 20803 (2016).
[41] X.-J. Wang et al., "Quantum Emitters with Narrow Band and High Debye–Waller Factor in Aluminum Nitride Written by Femtosecond Laser" Nano Letters **23** 2743 (2023).
[42] J. F. Ziegler, "SRIM-2003" Nuclear Instruments and Methods in Physics Research Section B: Beam Interactions with Materials and Atoms **219-220** 1027 (2004).
[43] Q. Ma, D. R. Clarke, "Optical fluorescence from chromium ions in sapphire: A probe of the image stress" Acta Metallurgica et Materialia **41** 1811 (1993).
[44] J. M. Binder et al., "Qudi: A modular python suite for experiment control and data processing" SoftwareX **6** 85 (2017).
[45] E. Corte et al., "Magnesium-Vacancy Optical Centers in Diamond" ACS Photonics **10** 101 (2023).